\documentclass[
 reprint,
 amsmath,amssymb,
 aps,
 prr,
 longbibliography,
 superscriptaddress
]{revtex4-2}

\usepackage{graphicx}
\usepackage{dcolumn}  
\usepackage{bm}       
\usepackage{physics}  
\usepackage{hyperref}
\usepackage{xcolor}
\usepackage{braket}
\usepackage{siunitx}
\usepackage{times}
\usepackage{microtype}
\usepackage{booktabs} 
\usepackage{float}
\usepackage{mathtools}
\usepackage{multirow}

\hypersetup{
    colorlinks = true,
    linkcolor = blue,
    citecolor = magenta,
    urlcolor  = teal
}

\begin{document}

\title{A Preparation Nonstationarity Loophole in Superconducting-Qubit Bell Tests}

\author{Prosanta Pal}
\affiliation{Department of Physics and Astronomy, Clemson University, Clemson, SC, USA}

\author{Shubhanshu Karoliya}
\affiliation{Indian Institute of Technology Mandi, Mandi, H.P. 175005, India}

\author{Gargee Sharma}
\affiliation{Indian Institute of Technology Delhi, Hauz Khas, New Delhi 110016, India}

\author{Ramakrishna Podila}
\affiliation{Department of Physics and Astronomy, Clemson University, Clemson, SC, USA}

\date{\today}


\begin{abstract}
Bell or Clauser–Horne–Shimony–Holt (CHSH) tests on superconducting quantum processors are commonly interpreted
under the assumption that repeated circuit executions sample a single,
stationary preparation ensemble.
Here we show that this assumption can be violated on contemporary hardware,
with direct implications for the interpretation of observed Bell violations.
We introduce an ensemble-divergence framework in which slow temporal drift of
the preparation process induces context-dependent effective ensembles, even
when measurement independence and locality are preserved.
This leads to a relaxed Bell bound $|S|\le 2+6\delta_{\mathrm{ens}}$, where
$\delta_{\mathrm{ens}}$ quantifies preparation nonstationarity. Because $\delta_{\mathrm{ens}}$ is not directly observable, we develop an
operational witness $\delta_{\mathrm{op}}$ based on bin-resolved outcome
statistics for fixed measurement channels.
Using Pauli-axis measurements on IBM superconducting processors, we observe
statistically significant operational drift that persists after full two-qubit readout mitigation, ruling out measurement artifacts.
In contrast, drift extracted from CHSH-optimal measurements is eliminated by
mitigation, demonstrating that such settings are unsuitable for diagnosing
preparation nonstationarity. We further show that the observed Bell violations imply only modest ensemble divergences, comparable in scale to those required in Hall-type measurement-dependence models, but arising here solely from preparation drift combined with experimental scheduling.
Our results identify a preparation-dependent loophole relevant to Bell tests
on noisy intermediate-scale quantum devices and highlight the necessity of
drift-aware protocols for reliable quantum certification.
\end{abstract}

\maketitle

\section{Introduction}

Superconducting-qubit processors have rapidly progressed from
proof-of-concept devices to mature experimental platforms for quantum
algorithms, noise characterization, and foundational tests of quantum
mechanics~\cite{Kjaergaard2020,Arute2019,Ansmann2009,Chow2014,Neill2018,
bharti2022noisy,tilly2022variational,fauseweh2024quantum,
bauer2023quantum,scholl2021quantum}.
Improvements in gate fidelity, scalable control electronics, and coherent
multi-qubit operation have positioned superconducting circuits as leading
candidates for both near-term quantum advantage demonstrations and precision
tests of quantum correlations~\cite{Barends2014,Storz2023}.
Operationally, experiments on these platforms consist of repeated executions
of a prepare--evolve--measure cycle, with each shot comprising state
preparation, coherent control, and projective measurement
~\cite{Koch2007,NielsenChuang}.
Most theoretical analyses and experimental interpretations implicitly assume
that these repetitions sample a single, stationary preparation process, often
modeled as an independent and identically distributed (i.i.d.) completely
positive trace-preserving (CPTP) map acting identically on every shot
~\cite{NielsenChuang}.
Noise-modeling frameworks such as Qiskit Aer adopt this assumption explicitly,
using backend-calibrated stationary CPTP channels to describe repeated
executions~\cite{javadi2024quantum}.

A growing body of experimental work, however, has established that this
i.i.d.\ assumption is frequently violated on superconducting hardware.
Observed deviations include slow temporal drift, context-dependent responses,
and correlated errors in state preparation and readout
~\cite{rudinger2019probing,Proctor2020,mcewen2021removing,
livingston2022experimental,hutchings2017tunable,kelly2015state,
wenner2011surface,howard2014contextuality,budroni2022kochen,
Kim2023,zhao2023mitigation,harper2020efficient}.
These effects arise primarily from low-frequency fluctuations in classical
control parameters, such as microwave amplitudes and phases, mixer imbalances,
dc offsets, flux noise, and residual qubit detunings, that introduce slowly
varying imperfections into the quantum process.
In multi-qubit architectures, shared-control crosstalk and calibration drift
can further induce context dependence, whereby nominally identical operations
exhibit small but systematic biases that depend on circuit structure or
measurement basis
~\cite{Kim2023,harper2020efficient,howard2014contextuality,
budroni2022kochen,zhao2023mitigation}.
As a result, repeated executions effectively sample from an ensemble of
slowly evolving preparation maps rather than a single stationary one, leading
to nonstationary outcome statistics
~\cite{Proctor2020,harper2020efficient,mcewen2021removing,
goswami2021experimental,filenga2020non,Setiawan2025}.

While nonstationarity and context dependence are well documented, their
implications for Bell-type tests on superconducting hardware remain
insufficiently explored.
In the standard Bell or Clauser--Horne--Shimony--Holt (CHSH) framework
~\cite{Clauser1969,ClauserHorne1974,Brunner2014},
the classical bound $|S|\le 2$ relies on two key assumptions: locality and a
single, context-independent hidden-variable distribution $\pi(\lambda)$
underlying all correlators.
Relaxing either assumption can increase the maximal attainable value of $S$.
Hall~\cite{hall2010local} formalized one such relaxation by allowing
correlations between the hidden variable and the measurement settings,
quantified by a measurement-dependence parameter
$M$. In this framework, modest measurement dependence
($M\gtrsim 14\%$) suffices for local models to reproduce the quantum value
$|S|=2\sqrt{2}$, yielding a relaxed bound $|S|\le 2+3M$, with subsequent
refinements
~\cite{friedman2019relaxed,barrett2011much,li2023tight,thinh2013bell}.
Hall’s parameter probes violations of measurement independence and is often
interpreted in terms of signaling or correlations between hidden variables and
setting choices.

In this work, we develop a different mechanism that relaxes the assumption of preparation stationarity. On superconducting processors, slow temporal variation of control parameters can cause the effective preparation ensemble to change over the duration of a Bell experiment. Because different CHSH settings may be executed at systematically different
times, the corresponding correlators can sample distinct preparation
ensembles even when the setting choices themselves remain free and random. We formalize this by introducing ensemble divergence $\delta_{\mathrm{ens}}$, by defining it to be the largest statistical difference between the preparation ensembles sampled by different Bell correlators (Eq.~\ref{eq:delta_ens_def}). Within local hidden-variable (LHV) models with bounded outcomes, nonzero
$\delta_{\mathrm{ens}}$ leads to a relaxed Bell inequality
$|S|\le 2+6\delta_{\mathrm{ens}}$, even in the absence of signaling or
measurement dependence. Because the underlying preparation ensemble is not an observable, preparation nonstationarity is assessed at the level of observable statistics. We introduce an operational drift parameter $\delta_{\mathrm{op}}$ which quantifies the maximal change in outcome distributions observed for a fixed measurement context across different temporal subsets of an experiment. A nonvanishing $\delta_\mathrm{op}$ signals the presence of preparation nonstationarity, i.e., $\delta_{\mathrm{ens}}>0$ within LHV models, even though it does not quantify the divergence between different contexts. In Bell–CHSH experiments, nonuniform execution schedules can convert such purely temporal drift into apparent context dependence by assigning different correlators to different temporal windows.



We experimentally demonstrate how preparation nonstationarity gives rise to apparent violations of the standard CHSH bound under realistic
execution conditions on IBM quantum computers. Our aim is not to survey or characterize this effect
across all qubit pairs, but to establish its existence as a distinct mechanism
by which Bell inequalities can be affected, independent of previously studied
relaxations such as measurement dependence. To this end, we perform CHSH
experiments with controlled scheduling protocols on IBM superconducting
quantum processors (\texttt{ibm\_fez} and \texttt{ibm\_torino}). We diagnose preparation nonstationarity using Pauli-axis measurements that
define fixed, time-independent classical channels, and compare the results
with matched Monte Carlo (MC) null models and noise-calibrated simulations.
We then examine how the independently observed preparation drift constrains
the interpretation of Bell violations obtained from CHSH-optimal measurements.
Our results demonstrate that Bell violations on contemporary superconducting
hardware can coexist with a breakdown of the preparation-stationarity premise,
opening a preparation-dependent loophole that is logically distinct from
Hall-type measurement dependence.

\section{Methods}
\label{sec:methods}

\subsection{Ensemble-conditioned CHSH bounds}
\label{sec:ensemble-bell}
In the standard Bell--CHSH framework~\cite{Bell1964,Clauser1969}, the four
measurement contexts
\((a,b)\in\{(x,y),(x',y),(x,y'),(x',y')\}\)
are assumed to probe a single, stationary preparation ensemble described by a
setting-independent hidden-variable distribution \(\pi(\lambda)\).
Together with measurement independence, the assumption that the setting choices
are statistically uncorrelated with \(\lambda\), this stationarity underlies the
derivation of the classical bound \(|S|\le 2\).

Hall~\cite{hall2010local} investigated a controlled relaxation of measurement
independence by allowing correlations between the hidden variable and the
measurement settings. In this approach, one permits correlations between the hidden variable
\(\lambda\) and the measurement settings \((a,b)\), so that the conditional
distribution \(p(\lambda|a,b)\) differs from the unconditional distribution
\(p(\lambda)\); equivalently, the joint distribution fails to factorize as
\(p(\lambda,a,b)=p(\lambda)p(a,b)\), or \(p(a,b|\lambda)\neq p(a,b)\). The degree of measurement
dependence is quantified by a parameter \(M\), defined as the maximum
$d_{TV}$ between the setting-conditioned distributions
\(p(\lambda|a,b)\), leading to the relaxed bound \(|S|\le 2+3M\).
Subsequent work has derived tight inequalities and information-theoretic
characterizations within this framework~\cite{friedman2019relaxed,barrett2011much,li2023tight,Larsson2014}.
These analyses exclusively concern correlations of the form \(p(a,b|\lambda)\).

The mechanism considered here relaxes a different assumption. We maintain
measurement independence in the operational sense,
\(p(a,b|\lambda)=p(a,b)\), and introduce no correlation between the hidden
variables and the setting choices. Instead, we relax the assumption of
preparation stationarity. In superconducting quantum processors, slow temporal
variation of control parameters can cause the effective preparation ensemble to
change over the duration of a Bell experiment. Since different setting pairs may
be executed at systematically different times, the corresponding correlators
can sample distinct preparation ensembles even when the setting choices remain
free and random.

Accordingly, the correlator for a given setting pair \((a,b)\) is described by
\begin{equation}
E_{ab}
=
\int d\lambda\, \pi_{ab}(\lambda)\,A(a,\lambda)B(b,\lambda),
\end{equation}
where \(\pi_{ab}(\lambda)\) denotes the hidden-variable distribution associated
with the preparation ensemble during the subset of runs in which \((a,b)\) is
implemented. To quantify deviations from ensemble stationarity, we define the
ensemble divergence
\begin{equation}
\delta_{\mathrm{ens}}
=
\max_{(a,b),(a',b')}
d_{\mathrm{TV}}\!\bigl(\pi_{ab},\pi_{a'b'}\bigr),
\label{eq:delta_ens_def}
\end{equation}
with \(d_{\mathrm{TV}}(\mu,\nu)=\tfrac{1}{2}\int d\lambda\,|\mu(\lambda)-\nu(\lambda)|\).

We stress that \(\pi_{ab}(\lambda)\) should not be interpreted as a conditional
distribution \(p(\lambda|a,b)\) in the sense of Hall \cite{hall2010local}. Rather, it represents an
effective description of preparation statistics associated with different
temporal subsets of the experiment. Consequently, the parameter
\(\delta_{\mathrm{ens}}\) quantifies a breakdown of preparation stationarity,
not a limitation of freedom of choice. It therefore relaxes a logically distinct
assumption of the CHSH derivation from Hall’s measurement-dependence parameter
\(M\), and the two quantities are not directly interchangeable.

The effect of
\(\delta_{\mathrm{ens}}\) on the Bell parameter can be bounded within a local
hidden-variable model with \(|A(a,\lambda)|,|B(b,\lambda)|\le 1\). Writing
\(X_{ab}(\lambda)\coloneqq A(a,\lambda)B(b,\lambda)\) and choosing a
reference ensemble \(\pi(\lambda)\equiv\pi_{xy}(\lambda)\), we decompose
\begin{align}
E_{ab}
&=
\int d\lambda\,\pi_{ab}(\lambda)\,X_{ab}(\lambda)
\label{eq:Eab_start}
\\[2pt]
&=
\int d\lambda\,\pi(\lambda)\,X_{ab}(\lambda)
+
\int d\lambda\,\bigl[\pi_{ab}(\lambda)-\pi(\lambda)\bigr]X_{ab}(\lambda)
\label{eq:Eab_decomp}
\\[2pt]
&\equiv
E_{ab}^{(0)} + \Delta E_{ab}.
\label{eq:Eab_split}
\end{align}
Inserting Eqs.~\eqref{eq:Eab_split} into
\(S = E_{xy} + E_{xy'} + E_{x'y} - E_{x'y'}\) and using
\(\Delta E_{xy}=0\) by construction yields \(S = S_0 + \Delta S\) with a
``reference'' contribution
\(S_0 = E_{xy}^{(0)} + E_{xy'}^{(0)} + E_{x'y}^{(0)} - E_{x'y'}^{(0)}\).
For each \(\lambda\) the deterministic combination
\(X_{xy} + X_{xy'} + X_{x'y} - X_{x'y'}\) has magnitude at most \(2\), so the
standard CHSH argument gives \(|S_0|\le 2\).

For the correction terms we use \(|X_{ab}(\lambda)|\le 1\) and the definition of
\(d_{\mathrm{TV}}\):
\begin{align}
|\Delta E_{ab}|
&=
\left|
\int d\lambda\,\bigl[\pi_{ab}(\lambda)-\pi(\lambda)\bigr]X_{ab}(\lambda)
\right|
\label{eq:DeltaE_start}
\\[2pt]
&\le
\int d\lambda\,\bigl|\pi_{ab}(\lambda)-\pi(\lambda)\bigr|
=
2\,d_{\mathrm{TV}}\!\bigl(\pi_{ab},\pi\bigr).
\label{eq:DeltaE_bound}
\end{align}
By Eq.~\eqref{eq:delta_ens_def},
\(d_{\mathrm{TV}}(\pi_{ab},\pi)\le\delta_{\mathrm{ens}}\) for all \((a,b)\), so
\(|\Delta E_{ab}|\le 2\,\delta_{\mathrm{ens}}\) for the three nonreference contexts
\((xy',x'y,x'y')\). Hence
\begin{equation}
|\Delta S|
\le
|\Delta E_{xy'}| + |\Delta E_{x'y}| + |\Delta E_{x'y'}|
\le
6\,\delta_{\mathrm{ens}},
\end{equation}
and we obtain the relaxed CHSH inequality
\begin{equation}
|S|
\le
|S_0| + |\Delta S|
\le
2 + 6\,\delta_{\mathrm{ens}}.
\label{eq:Sr_relaxed_main}
\end{equation}
Equation~\eqref{eq:Sr_relaxed_main} shows that the variation of preparation ensembles alone can increase the maximal Bell parameter by an
amount proportional to \(\delta_{\mathrm{ens}}\), without relaxing locality or
freedom-of-choice assumptions. In particular, a violation \(|S|>2\) implies
\(\delta_{\mathrm{ens}}\ge(|S|-2)/6\) for any local hidden-variable model
with deterministic or bounded outcomes.

\subsection{Operational ensemble drift}
The ensemble divergence \(\delta_{\mathrm{ens}}\) defined in Eq.~\eqref{eq:delta_ens_def}
characterizes variability of the hidden preparation ensembles
\(\pi_{ab}(\lambda)\) across temporal bins used in experiments. Because \(\pi_{ab}\) is not directly
observable, we assess nonstationarity at the level of the operational outcome
statistics. For a fixed context \((a,b)\), let \(p^{(i)}_{ab}(x)\) denote the
empirical distribution over bitstrings in time bin \(i\). We define the
operational ensemble drift
\begin{equation}
\delta_{\mathrm{op}}(ab)
\;=\;
\max_{i<j}
d_{\mathrm{TV}}\!\bigl(p^{(i)}_{ab},p^{(j)}_{ab}\bigr),
\qquad
\delta_{\mathrm{op}}^{global}
\;=\;
\max_{(a,b)} \delta_{\mathrm{op}}(ab),
\label{eq:deltaop_def}
\end{equation}
which quantifies the maximal variation of the observed statistics for a single
measurement channel. Since each context induces a fixed classical map
\(\pi_{ab}\mapsto p_{ab}\), the contractivity of TV distance implies
\begin{equation}
\delta_{\mathrm{op}}(ab)
\;\le\;
d_{\mathrm{TV}}\!\bigl(\pi_{ab}^{(i)},\pi_{ab}^{(j)}\bigr)
\;\le\;
\delta_{\mathrm{ens}},
\label{eq:deltaop_le_delta}
\end{equation}
so any nonzero operational drift certifies \(\delta_{\mathrm{ens}}>0\). However,
because different CHSH settings correspond to distinct measurement channels,
Eq.~\eqref{eq:deltaop_le_delta} does not extend across settings.
Consequently, \(\delta_{\mathrm{op}}\) provides a valid witness of
preparation nonstationarity but does not quantify the magnitude of
\(\delta_{\mathrm{ens}}\) or its variation between contexts. 

An additional operational issue concerns the choice of measurement bases used to evaluate \(\delta_{\mathrm{op}}\). While CHSH-optimal settings (measurement axes in the \(XZ\) plane separated by angles \(\{0,\tfrac{\pi}{2}\}\) on one qubit and \(\{\pm\tfrac{\pi}{4}\}\) on the other) maximize the Bell violation \cite{lima2011optimal, gill2023optimal}, they correspond to non-orthogonal, continuously calibrated measurement axes. Slow drift in control amplitudes, phases, or qubit frequencies can therefore induce basis-dependent mixing of outcome probabilities, making it difficult to distinguish genuine preparation nonstationarity from measurement-axis drift at the level of raw statistics \cite{rosset2012imperfect}. For this reason, we evaluate \(\delta_{\mathrm{op}}\) using fixed Pauli \(X,Y,Z\)
measurement axes. Pauli measurements define orthogonal, time-independent
classical channels whose outcome probabilities are directly interpretable and
stable under small coherent control errors. This choice aligns with IBM Quantum calibration protocols, in which Pauli measurements are routinely
calibrated and monitored as fixed reference bases. Consequently, variations of \(p^{(i)}_{ab}\) across temporal bins can be attributed conservatively to changes in the effective preparation ensemble rather than to measurement-channel drift.

\subsection{Qiskit-Aer simulations with real backend noise}
\label{sub:qiskit}

To select an appropriate temporal binning for the hardware experiments, we
performed forward simulations using \texttt{Qiskit Aer} with noise models
calibrated from each backend at the time of data acquisition. These simulations
are not intended to reproduce the experimental data quantitatively, but rather
to assess how operational measures of preparation nonstationarity respond to
realistic levels of slow drift under controlled conditions.

\paragraph*{Measurement settings.}
In these simulations, we implement the CHSH measurement contexts by performing
projective measurements of each qubit in the singlet state along specified Pauli axes. The four
contexts \((X,Y)\), \((X,Z)\), \((Z,Y)\), and \((Z,Z)\) correspond to joint
measurements of the observables \(\sigma_X^A\otimes\sigma_Y^B\),
\(\sigma_X^A\otimes\sigma_Z^B\), \(\sigma_Z^A\otimes\sigma_Y^B\), and
\(\sigma_Z^A\otimes\sigma_Z^B\), respectively. Operationally, these measurements
are implemented by applying single-qubit rotations that map the chosen Pauli
basis onto the computational basis, followed by standard projective measurement
in the \(Z\) basis on both qubits. As discussed earlier, these settings are chosen to avoid conflating drift-induced nonstationarity with angle-dependent sensitivity amplification that arises for optimal CHSH settings. 

\paragraph*{Drift model.}
To emulate experimentally realistic nonstationarity, we augment the real backend Aer
noise model with a weak, coherent drift applied at the state-preparation level.
Specifically, for each temporal bin \(b\in\{1,\dots,B\}\) and each CHSH context,
both qubits undergo small phase rotations
\begin{equation}
U_{\mathrm{drift}}^{(b)}
=
\exp\!\left[-i\,\theta_b Z_a\right]
\exp\!\left[+i\,\theta_b Z_b\right],
\end{equation}
where \(\theta_b\) varies smoothly across the acquisition window.
In the simulations presented here, \(\theta_b\) is swept linearly from
\(-\theta_{\max}\) to \(+\theta_{\max}\) across all bins and contexts, mimicking
slow calibration drift or control-parameter wander on timescales longer than a
single bin but shorter than the full experiment.
We consider \(\theta_{\max}\in\{0,10^{-2},10^{-1}\}\), corresponding to
bin-to-bin probability variations ranging from purely statistical fluctuations
to percent-level changes comparable to those observed in preliminary hardware
runs.

\paragraph*{Scheduling protocols.}
Two execution schedules are simulated: an interleaved (round-robin) schedule,
where all four contexts are sampled within each bin, and an unbalanced
(blocked) schedule, where all bins of a given context are executed contiguously.
Both schedules use identical total shot counts (1024 shots per bin) and drift profiles, differing
only in how drift is mapped onto measurement contexts. This allows us to
disentangle drift-induced ensemble variation from trivial shot-count effects.

\paragraph*{Bin-number scan and null model.}
For each backend and drift strength, we scan the number of bins
\(B\in\{3,6,9,12\}\) while keeping the number of shots \emph{per bin} fixed at
\(1024\). As a result, the total number of shots per context scales linearly with
\(B\).
For each configuration, we compute the operational ensemble divergence
\(\delta_{\mathrm{op}}\) as the maximum $d_{TV}$
between empirical outcome distributions drawn from different bins.
To assess whether the observed \(\delta_{\mathrm{op}}\) exceeds
what is expected from finite sampling alone, we construct a MC null
distribution by resampling i.i.d.\ multinomial data with the same number
of bins and shots per bin as in the simulated experiment. Importantly, because the shots per bin are held fixed, increasing \(B\) does not
simply sharpen statistical resolution; it also inflates the null distribution of
\(\delta_{\mathrm{op}}\) through extreme-value statistics. As a
result, the comparison between observed and null values should not be interpreted
as a hypothesis test in the conventional sense. Rather, it provides an
operational diagnostic of the regime in which drift-induced ensemble variation
becomes distinguishable from finite-sample fluctuations under experimentally
realistic constraints.

\subsection{Hardware experiments on superconducting quantum processors}

\paragraph*{CHSH measurements along Pauli axes.}Hardware experiments are performed using the same binning, scheduling, and statistical analysis protocols as those established in the Qiskit-Aer simulations (Sec. \ref{sub:qiskit}),
with the sole difference that all circuits are executed on real IBM
superconducting quantum processors and no explicit drift is introduced. We prepared the singlet state 
\(|\Psi^{-}\rangle=(|01\rangle-|10\rangle)/\sqrt{2}\) 
on connected qubit pairs ((40, 41) on \texttt{ibm\_torino} and (0, 1) on \texttt{ibm\_fez})  using standard Hadamard–CNOT sequences followed by local phase adjustments.

Our objective is to demonstrate the existence of a preparation
nonstationarity loophole under realistic experimental conditions, rather than to establish its prevalence across all qubit pairs or devices. Accordingly, we do not attempt an exhaustive survey. We performed CHSH tests on multiple candidate qubit pairs across the two backends ($(0,1)$, $(2,3)$, and $(40,41)$ on \texttt{ibm\_torino} and $(0,1)$, $(25,37)$, $(34, 35)$, $(37, 45)$, and $(140,141)$ on \texttt{ibm\_fez}), but restricted the more time-intensive $\delta_{\mathrm{op}}$ measurements to two representative
pairs due to the substantial cost ($\sim$96\,s wall-clock time per pair
for combined round-robin and unbalanced schedules with $B=6$ bins). These candidate pairs were screened a priori using publicly available
calibration data (including readout error, two-qubit gate error, and
$T_1/T_2$ variability) together with short pilot runs to identify pairs with
elevated temporal variability. 

On all hardware runs, data acquisition is divided into $B=6$ temporal bins with
$1024$ shots per bin per context, and both round-robin and unbalanced execution
schedules are employed to map temporal variation onto measurement contexts.
Measurements are performed using fixed Pauli axes $(X,Y)$, $(X,Z)$, $(Z,Y)$,
and $(Z,Z)$, implemented via single-qubit basis rotations followed by
projective measurement in the computational basis. While Pauli measurements are
often treated operationally as defining time-independent classical channels,
we note that on superconducting qubits they are realized through pre-rotations
and therefore inherit any residual instability in single-qubit control \cite{gao2021practical, ware2021experimental}. This
effect must be distinguished from preparation nonstationarity and explicitly
controlled. As discussed later in Sec. \ref{sub:deltaopexp}, we address this by bounding the stability of
the relevant single-qubit rotations over the acquisition window and by
verifying that their contribution is parametrically smaller than the observed
bin-to-bin variations. With this control in place, the use of Pauli axes avoids conflating preparation
nonstationarity with the basis-dependent sensitivity inherent to CHSH-optimal
settings. Operational ensemble drift $\delta_{\mathrm{op}}$ is computed from
the bin-resolved outcome distributions exactly as in the simulations, and its
statistical significance is assessed using a MC null constructed with
the same number of bins and shots per bin.

\paragraph*{CHSH measurements along optimal axes.}
In addition to Pauli-axis measurements used to diagnose preparation
nonstationarity, we perform standard CHSH tests using measurement settings that
maximize the quantum violation for a singlet state.
Specifically, each qubit is measured along axes in the \(X\!-\!Z\) plane defined
by angles \(\theta\) with respect to \(Z\), with settings
\(\theta_A\in\{0,\pi/2\}\) for qubit~A and
\(\theta_B\in\{\pm\pi/4\}\) for qubit~B.
These correspond to the canonical CHSH-optimal choices yielding
\(|S|=2\sqrt{2}\) for an ideal singlet.
Operationally, a measurement along angle \(\theta\) is implemented by a single
\(R_Y(-\theta)\) rotation followed by projective measurement in the
computational basis. For each qubit pair and backend ((40, 41) on \texttt{ibm\_torino} and (0, 1) on \texttt{ibm\_fez}), we prepare the singlet state
\(|\psi^-\rangle=(|01\rangle-|10\rangle)/\sqrt{2}\) and evaluate the four
correlators entering the CHSH parameter \(S\).
Two scheduling protocols are employed: a round-robin schedule, in which all four
settings are executed within each repeat, and an unbalanced (or blocked) schedule, in which each
setting is executed contiguously across repeats.
Each correlator is estimated from \(1024\) shots per circuit, and \(S\) is
computed solely from counts aggregated over all repeats for a given setting. We emphasize that these CHSH-optimal measurements are used exclusively to
estimate Bell violations.
Because the corresponding measurement channels depend on continuously calibrated
analog rotation angles, they are not used to extract
\(\delta_{\mathrm{op}}\), as slow basis drift can mimic preparation
nonstationarity at the level of outcome statistics.

\paragraph*{Two-qubit readout mitigation.}
To reduce state-preparation-and-measurement (SPAM) bias in both the aggregated
CHSH estimator and the bin-resolved drift statistics, we apply classical
measurement-error mitigation based on an empirically calibrated two-qubit
assignment matrix $M\in\mathbb{R}^{4\times4}$. Here SPAM refers to deviations
between the intended preparation/measurement procedure and the realized
classical readout statistics, which can systematically distort the observed
outcome frequencies. For each backend and physical qubit pair, the matrix elements are defined as
$M_{x,y}=P(\mathrm{meas}=x\mid \mathrm{prep}=y)$ with $x,y\in\{00,01,10,11\}$,
and are obtained by preparing each computational basis state $|y\rangle$ and
estimating the resulting outcome distribution over $x$. Calibrating a single
joint channel on the two-qubit outcome space captures correlated assignment
effects and leakage between outcomes that may not be represented by
single-qubit figures of merit or separable models--a point emphasized by
recent analyses of multiplexed readout quality beyond assignment
fidelity~\cite{di2025benchmarking}. Mitigated distributions are obtained by linear
inversion, $\hat p^{\,\mathrm{mit}}=M^{-1}\hat p$, with conditioning checks and
renormalization to the probability simplex to limit noise amplification \cite{Bravyi2021, Endo2018, maciejewski2020mitigation}.  Unlike linear inversion, more sophisticated mitigation strategies based on Bayesian
inference or full POVM reconstruction can incorporate additional assumptions
or priors that may adapt to statistical fluctuations and partially absorb
time-dependent effects into the inference itself~\cite{cosco2024bayesian, maciejewski2020mitigation}.
Such adaptivity, while advantageous for state estimation, would obscure the
drift signatures we seek to quantify. For this reason, we adopt linear
inversion with standard numerical safeguards as a conservative and
operationally well-defined procedure for isolating preparation nonstationarity. Importantly, we apply the same mitigation procedure independently to each temporal bin prior to computing $\delta_{\mathrm{op}}$, so that the drift
diagnostic compares bin-to-bin variations in inferred pre-assignment
statistics rather than raw readout artifacts. 

\section{Results}
\label{sec:results}

\subsection{Operational ensemble drift along Pauli axes}
\label{sub:deltaopexp}
\begin{figure}[]
    \centering
    \includegraphics[width=\columnwidth]{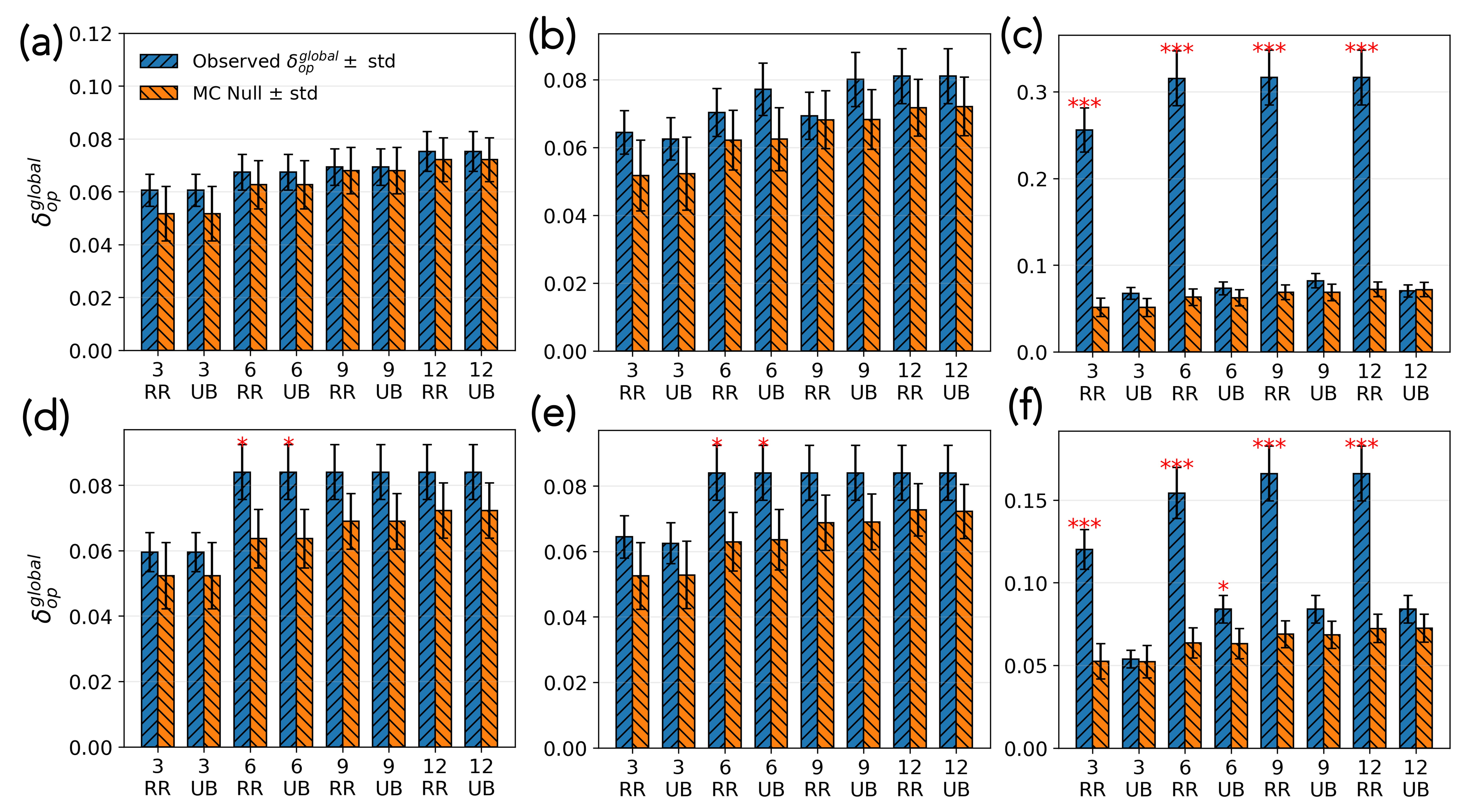}
    \caption{
    Operational ensemble drift $\delta_{op}^{global}$ extracted from Monte-Carlo–simulated CHSH experiments under controlled preparation drift.
    Panels (a)–(c) correspond to the \texttt{ibm\_fez} backend using qubit pair (0,1), while panels (d)–(f) correspond to the \texttt{ibm\_torino} backend using qubit pair (40,41).
    Vertically aligned panels share the same drift amplitude: (a,d) $\theta_{\max}=0$, (b,e) $\theta_{\max}=0.01$, and (c,f) $\theta_{\max}=0.1$.
    For each panel, results are shown as a function of the number of temporal bins $B$ and scheduling protocol (RR: round-robin; UB: unbalanced).
    Blue hatched bars denote the observed $\delta_{op}^{global}$ $\pm$ one standard deviation, while orange hatched bars indicate the Monte-Carlo null mean $\pm$ one standard deviation.
    Red asterisks indicate statistically significant deviations from the null distribution (* $p<0.05$, ** $p<0.01$, *** $p<10^{-3}$).
    }
    \label{fig:deltaop_global_drift}
\end{figure}

Figure~\ref{fig:deltaop_global_drift} shows the global operational drift
\(\delta_{\mathrm{op}}^{\mathrm{global}}\) [Eq.~\eqref{eq:deltaop_def}] extracted from Pauli-basis measurements
in noise-calibrated simulations as a function of the number of temporal bins
\(B\), execution schedule, and imposed drift amplitude, together with a matched MC null. In the absence of drift (\(\theta_{\max}=0\)), the observed
\(\delta_{\mathrm{op}}^{\mathrm{global}}\) follows the null for all \(B\), with the weak increase reflecting extreme-value broadening rather than physical nonstationarity. At intermediate drift (\(\theta_{\max}=10^{-2}\)), deviations from the null become marginally resolvable, with the first consistent separation appearing at \(B\approx6\), particularly under unbalanced scheduling. For larger drift (\(\theta_{\max}=10^{-1}\)), the observed \(\delta_{\mathrm{op}}\) saturates for \(B\gtrsim6\), indicating diminishing returns from further binning while the null continues to broaden.
Thus, \(B=6\) represents the minimal binning that resolves drift-induced
nonstationarity without incurring unnecessary statistical inflation or shot
overhead. Identifying this minimal binning is operationally important.
On real IBM QPUs, increasing \(B\) increases total circuit executions without
improving sensitivity, directly inflating experimental cost.
Moreover, larger \(B\) increases the likelihood that bins are executed at widely separated wall-clock times due to queueing and priority scheduling, introducingadditional uncontrolled temporal variation.
Thus, \(B=6\) represents the smallest binning that resolves drift-induced
nonstationarity while minimizing QPU overhead and avoiding artificial temporal
segregation.
All hardware experiments therefore use \(B=6\).

\begin{figure}[]
    \centering
    \includegraphics[width=0.8\columnwidth]{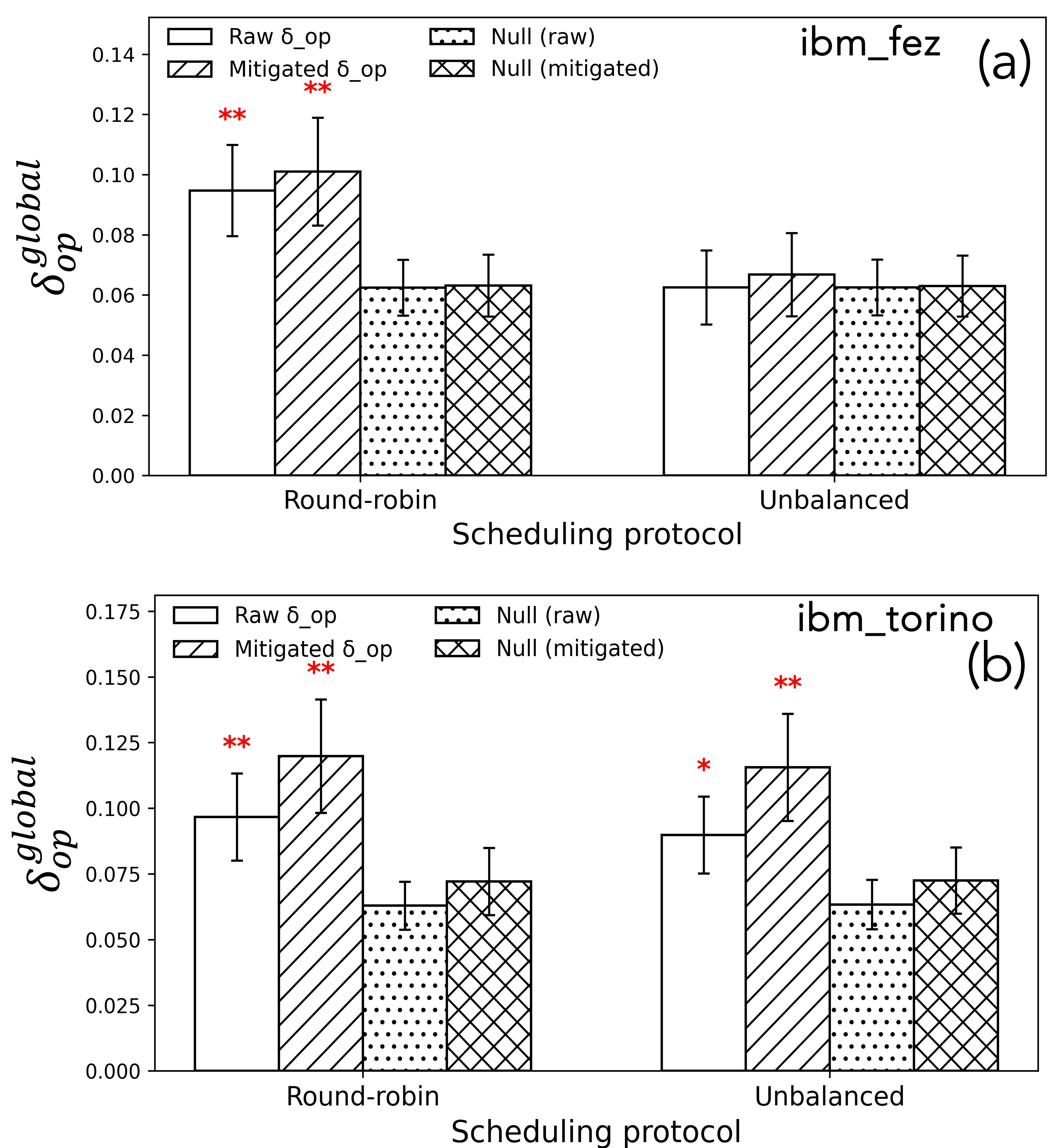}
    \caption{
    Experimentally extracted operational ensemble drift $\delta_{\mathrm{op}}$ on (a) the \texttt{ibm\_fez} backend using qubit pair (0,1) and (b) the \texttt{ibm\_torino} backend using qubit pair (40,41). Results are shown for round-robin and unbalanced scheduling protocols. White and hatched blue bars denote the observed raw and readout-mitigated $\delta_{\mathrm{op}}$, respectively, while dotted and cross-hatched bars indicate the corresponding Monte-Carlo null expectations constructed using identical shot budgets and scheduling structure. Error bars denote one standard deviation.
    Red asterisks indicate statistically significant deviations from the null (* $p<0.05$, ** $p<0.01$).
    }
    \label{fig:deltaop_exp}

\end{figure}

Figure ~\ref{fig:deltaop_exp} summarizes the experimentally observed operational ensemble drift $\delta_{\mathrm{op}}$ extracted from CHSH experiments executed on two distinct IBM superconducting quantum processors.
For each device, we compare round-robin and unbalanced scheduling protocols, reporting both raw and readout-mitigated estimates of $\delta_{\mathrm{op}}$, along with MC null distributions that incorporate finite-shot effects and the precise experimental scheduling structure. 

The simulations (Fig. \ref{fig:deltaop_global_drift}) establish a clear baseline expectation.
In the absence of preparation nonstationarity, the simulated null distributions predict $\delta_{\mathrm{op}}$ values that remain confined to a narrow range set by finite-shot fluctuations, with no systematic dependence on scheduling protocol.
This behavior is reflected in the null bars shown in Fig. \ref{fig:deltaop_exp}, which are comparable for round-robin and unbalanced execution. For both \texttt{ibm\_fez} and \texttt{ibm\_torino}, the observed $\delta_{\mathrm{op}}$ in round-robin exceeds the simulated null by multiple standard deviations, with statistical significance ranging from $p<0.05$ to $p<10^{-3}$.  The persistence of this excess after readout mitigation rules out classical measurement errors as the dominant mechanism: if readout errors were responsible, the mitigated experimental values would collapse toward the simulated null, which is not observed.
Under unbalanced scheduling, the experimentally observed $\delta_{\mathrm{op}}$ is substantially reduced for \texttt{ibm\_fez} and becomes consistent with the simulated null, in quantitative agreement with simulation predictions. However, \texttt{ibm\_torino} shows a significant difference even for unbalanced scheduling, suggesting stronger nonstationarity in preparation.

To further assess whether readout mitigation could artificially induce
schedule-dependent effects, we examined the conditioning of the two-qubit
assignment matrix $A$. For both \texttt{ibm\_fez} and \texttt{ibm\_torino},
$\kappa(A)$ is moderate ($\kappa\!\approx\!20$--24) and, for a given backend,
indistinguishable between round-robin and unbalanced executions with identical
binning and shot counts. Although small run-to-run fluctuations are observed,
as expected from routine recalibration, the conditioning shows no systematic
dependence on scheduling. To test the stability of readout mitigation over different bins with in each schedule, we
measured $M$ at four time points separated by six-bin blocks ($\sim$48\,s). The condition number remains stable, $\kappa_2(M)\approx 20$–24 across all snapshots, while
the change relative to the earliest calibration is small
($\|M_{\mathrm{early}}-M\|_F\approx 0.02$ and
$\max_{ij}|\Delta M_{ij}|\lesssim 0.02$). These variations are well within the
numerical stability of the inversion and show no evidence of time-dependent
degradation on the scale of the experiment. This indicates that readout-matrix inversion remains
numerically stable throughout the experiments and cannot account for the
observed schedule-dependent behavior in the CHSH results.

Pauli-axis $\delta_{\mathrm{op}}$ uses fixed single-qubit rotations
($\sqrt{X}/X$ and virtual $R_z$). Calibration data nearest the runs bound
single-qubit control drift to induce $\lesssim10^{-3}$ changes in outcome
probabilities, over an order of magnitude below the observed
$\delta_{\mathrm{op}}\approx0.06$–$0.09$ under unbalanced execution
(i.e., a separation of $\gtrsim 50\times$; ). The absence of schedule dependence in these calibration metrics rules out single-qubit rotation instability as the origin of the observed
$\delta_{\mathrm{op}}$.

The statistically significant operational nonstationarity observed in
Fig. \ref{fig:deltaop_exp} has direct implications for the interpretation of Bell tests on
superconducting hardware. As shown in Sec.~II~A, any local hidden-variable
model with ensemble divergence $\delta_{\mathrm{ens}}>0$ obeys a relaxed
bound $|S|\le 2+6\delta_{\mathrm{ens}}$. While $\delta_{\mathrm{ens}}$ is not
directly accessible, the experimentally observed $\delta_{\mathrm{op}}>0$
constitutes a verifiable witness that the assumption of a single stationary
preparation ensemble, required for the standard CHSH bound, is violated.
Accordingly, in regimes where $\delta_{\mathrm{op}}$ exceeds the finite-shot
null, the strict classical bound $|S|\le 2$ is no longer guaranteed by the
operational premises of the experiment, even in the absence of signaling or
freedom-of-choice violations. 

We emphasize that $\delta_{\mathrm{op}}$, extracted from fixed Pauli-axis
measurements, cannot be substituted into Eq.~\ref{eq:Sr_relaxed_main} to quantify Bell-bound
relaxation for CHSH-optimal settings, as different contexts define distinct
classical channels acting on the preparation ensemble. Accordingly, our
results do not claim a quantitative explanation of the observed CHSH values
in terms of Pauli-axis drift alone. Rather, the data in Fig. ~\ref{fig:deltaop_exp} establish a preparation-dependent loophole that is
logically distinct from Hall-type measurement dependence: measurement
independence is preserved, while temporal variation of the preparation
ensemble sampled by different correlators relaxes the assumptions required
for the standard CHSH bound. The persistence of $\delta_{\mathrm{op}}$ after
two-qubit readout mitigation and its sensitivity to scheduling protocol
indicate that this effect originates in state preparation rather than in
measurement artifacts.

\subsection{Operational tests along optimal CHSH settings}

\begin{figure}[]
    \centering
    \includegraphics[width=\linewidth]{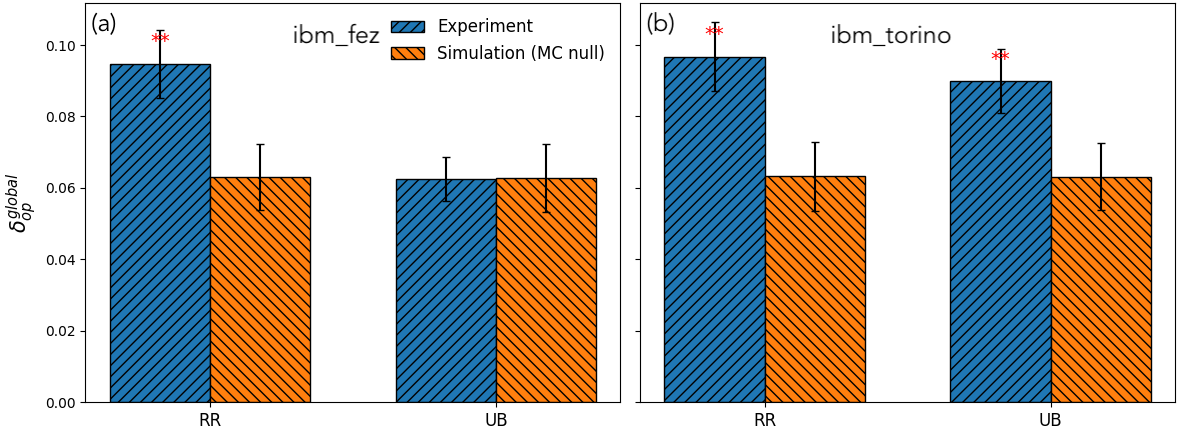}
    \caption{
    Operational drift $\delta_{\mathrm{op}}$ extracted from outcome statistics
    measured along CHSH-optimal axes.
    Panels (a,b) show raw experimental values for \texttt{ibm\_fez} (qubit pair
    (0,1)) and \texttt{ibm\_torino} (qubit pair (40,41)), respectively, compared
    with matched Monte Carlo (MC) null distributions for round-robin (RR) and
    unbalanced (UB) scheduling. While the raw data exhibit apparent deviations from the null, these
    differences are eliminated by mitigation (data not shown), rendering the experimental
    $\delta_{\mathrm{op}}$ statistically consistent with the null.
    This behavior indicates that $\delta_{\mathrm{op}}$ extracted from
    CHSH-optimal settings is dominated by basis-dependent measurement
    instabilities rather than genuine preparation nonstationarity.
    }
    \label{fig:deltaop_chsh_optimal_raw}
\end{figure}

For completeness, we also evaluate the operational drift $\delta_{\mathrm{op}}$
using outcome statistics obtained from CHSH-optimal measurement settings.
In the raw data Fig. \ref{fig:deltaop_chsh_optimal_raw} , the experimentally extracted $\delta_{\mathrm{op}}$
appears significantly larger than the corresponding MC null for both
devices and scheduling protocols. However, after applying full two-qubit
readout mitigation, the experimental values become statistically
indistinguishable from the null within uncertainty. This behavior contrasts
sharply with the Pauli-axis results of Fig.~\ref{fig:deltaop_exp}, where significant deviations persist after mitigation.

This sensitivity to mitigation reflects the fact that CHSH-optimal measurements are implemented via continuously calibrated analog rotations. Slow drift in control amplitudes, phases, or qubit frequencies can therefore induce basis-dependent mixing of outcome probabilities that mimics nonstationarity at the level of raw statistics. As a result, $\delta_{\mathrm{op}}$ extracted from CHSH-optimal settings conflates preparation variation with measurement-axis instability and does not provide a reliable witness of ensemble nonstationarity. For this reason, CHSH-optimal measurements are used here exclusively to estimate the Bell parameter $S$, while operational drift is diagnosed using fixed Pauli measurement axes.

\begin{figure}[]
    \centering
    \includegraphics[width=\linewidth]{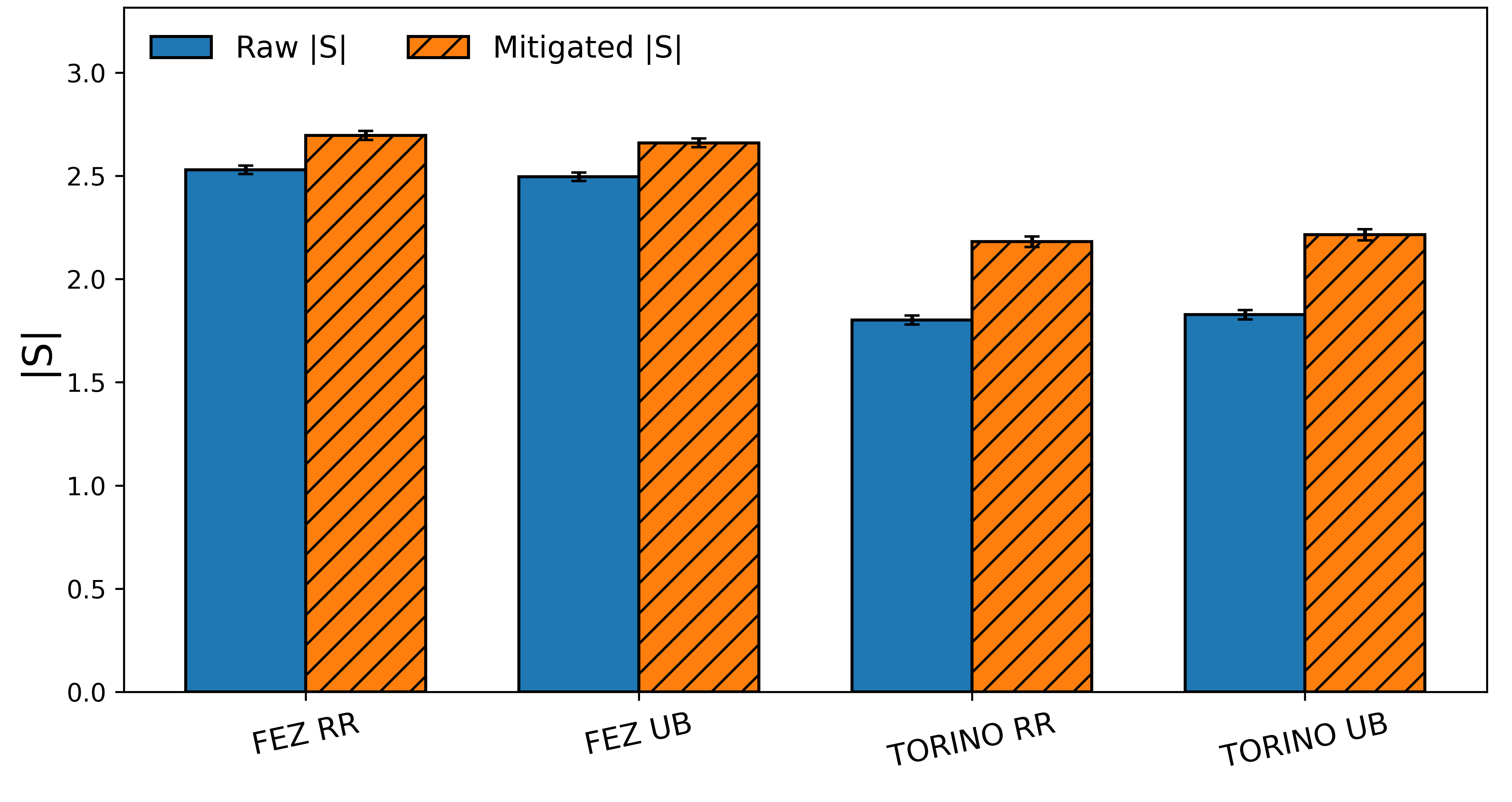}
    \caption{
    Absolute Bell parameter $|S|$ measured using CHSH-optimal settings on
\texttt{ibm\_fez} (qubit pair (0,1)) and \texttt{ibm\_torino} (qubit pair (40,41)).
Blue bars are computed from raw counts and hatched orange bars from the
corresponding readout-mitigated counts, shown for round-robin (RR) and
unbalanced (UB) schedules; error bars denote one standard deviation.
This figure reports the standard CHSH estimator of $S$ but does not test the
preparation-stationarity assumption required for the ideal bound $|S|\le 2$;
stationarity is assessed independently via the bin-resolved drift diagnostics
in Fig.~2. Consequently, $|S|$ should be interpreted jointly with Fig.~2 when
assessing whether the operational premises of the standard CHSH derivation are
satisfied on hardware.
    }
    \label{fig:chsh_optimal_absS}
\end{figure}

Figure~\ref{fig:chsh_optimal_absS} shows the experimentally measured Bell
parameter $|S|$ obtained using CHSH-optimal measurement settings.
As expected, readout mitigation systematically increases $|S|$ for both
devices and scheduling protocols, recovering violations of the classical
bound in cases where raw data lie below $|S|=2$.
Viewed in isolation, these results would be interpreted as evidence of
nonclassical correlations.

However, these CHSH measurements must be interpreted in light of the
independently observed preparation nonstationarity reported in Fig.~2.
The CHSH parameter $S$ is estimated from statistics aggregated over all
repetitions of a given measurement setting, and therefore probes only
time-averaged correlators.
As a result, CHSH-optimal measurements are insensitive to the temporal
structure of the experiment and cannot diagnose whether different
correlators sample the same underlying preparation ensemble.

Consequently, the observation of $|S|>2$ does not, by itself, guarantee that
the assumptions required for the standard Bell--CHSH bound are satisfied.
In particular, the Pauli-axis diagnostics establish that the effective
preparation ensemble varies across the experiment for \texttt{ibm\_fez}
(round-robin) and \texttt{ibm\_torino} (both schedules), implying
$\delta_{\mathrm{ens}}>0$.
In this regime, the relevant classical benchmark is no longer $|S|\le 2$,
but the relaxed bound $|S|\le 2+6\delta_{\mathrm{ens}}$, even though
$\delta_{\mathrm{ens}}$ itself cannot be inferred from CHSH-optimal data.

These results therefore demonstrate a preparation-dependent loophole that
persists even when measurement independence is operationally satisfied and
readout errors are mitigated. Bell violations observed under such conditions cannot be interpreted as
device-independent evidence against local realism without an accompanying
assessment of preparation stationarity. 

\begin{table}[]
\centering
\caption{No-signaling (marginal-consistency) summary for each device and
schedule. For each dataset, four marginal-comparison tests (two for Alice
and two for Bob) are performed. We report the maximum absolute deviation
$\max|\Delta P|$, the minimum two-sided $p$-value $\min p$, and the
Bonferroni-corrected minimum $p$-value
$\min p_{\mathrm{Bonf}}=\min(1,4\,\min p)$. Results are shown for raw counts
and, where indicated, for readout-mitigated distributions.}
\label{tab:nosignaling}
\vspace{0.25em}
\begin{tabular}{l l r r r}
\hline
Device / schedule & Version & $\max|\Delta P|$ & $\min p$ & $\min p_{\mathrm{Bonf}}$ \\
\hline
ibm\_fez (q0--1), RR & raw       & 0.0067 & 0.45 & 1 \\
ibm\_fez (q0--1), RR & mitigated & 0.0093 & 0.32 & 1 \\
ibm\_fez (q0--1), UB & raw       & 0.0179 & 0.047 & 0.19 \\
ibm\_fez (q0--1), UB & mitigated & 0.0116 & 0.19 & 0.79 \\
ibm\_torino (q40--41), RR & raw       & 0.0123 & 0.17 & 0.71 \\
ibm\_torino (q40--41), RR & mitigated & 0.0141 & 0.1 & 0.43 \\
ibm\_torino (q40--41), UB  & raw       & 0.0168 & 0.024 & 0.09 \\
ibm\_torino (q40--41), UB & mitigated & 0.1723 & $\ll 10^{-6}$ & $\ll 10^{-6}$ \\
\hline
\end{tabular}
\end{table}

As an operational consistency check, we examined marginal (no-signaling)
constraints in the CHSH data. For each context $(x,y)\in\{0,1\}^2$, we form
the one-party marginals
$P_A(a|x,y)=\sum_b P(a,b|x,y)$ and
$P_B(b|x,y)=\sum_a P(a,b|x,y)$ from the observed two-bit frequencies.
Standard no-signaling requires $P_A(a|x,y)=P_A(a|x,y')$ for fixed $x$ and
$P_B(b|x,y)=P_B(b|x',y)$ for fixed $y$. We test the four corresponding
comparisons using two-proportion $z$-tests and report the maximum absolute
marginal deviation $\max|\Delta P|$, the minimum two-sided $p$-value, and a
Bonferroni-corrected minimum $p$-value (factor $4$); results are summarized
in Table~\ref{tab:nosignaling}. Such empirical marginal tests have recently
been used as operational diagnostics in superconducting-qubit Bell
experiments~\cite{tabia2025almost}.

Crucially, these marginal-consistency tests implicitly assume stationarity
of the underlying preparation and measurement processes. While this
assumption underlies standard interpretations of no-signaling
diagnostics~\cite{tabia2025almost}, it must be reconsidered in the presence of temporal drift. In this case, the experimentally accessible probabilities
are time-averaged mixtures whose effective ensembles depend on the execution
schedule. Apparent violations of marginal consistency in aggregated data
therefore diagnose context-dependent sampling of a nonstationary ensemble,
rather than signaling or nonlocal influence. Consistent with this
interpretation, round-robin schedules, which enforce more uniform temporal
sampling, exhibit suppressed marginal deviations, while unbalanced schedules
show enhanced deviations. This behavior is fully compatible with
instantaneous locality at the level of individual executions and does not
contradict the $\delta_{\mathrm{op}}$ analysis, which is specifically
designed to quantify operational nonstationarity across temporal bins.

\subsection{Schedule-aware lower bound on $\delta_{\mathrm{ens}}$}

To connect the observed Bell violations to the preparation-nonstationarity
framework without identifying $\delta_{\mathrm{op}}$ with $\delta_{\mathrm{ens}}$,
we compute the minimal ensemble divergence required for a local model to
accommodate a measured value of $|S|$ using Eq.~\ref{eq:Sr_relaxed_main}:
$\delta_{\mathrm{ens}}^{\min}=(|S|-2)/6$.
From Fig.~\ref{fig:chsh_optimal_absS}, the mitigated CHSH values imply
$\delta_{\mathrm{ens}}^{\min}\approx 0.11$--$0.12$ for \texttt{ibm\_fez} and
$\delta_{\mathrm{ens}}^{\min}\approx 0.03$--$0.04$ for \texttt{ibm\_torino}.
These numbers should be interpreted only as the scale of preparation divergence
that would be necessary for a local explanation within our relaxed-bound
analysis; they do not follow from $\delta_{\mathrm{op}}$, which is channel
dependent and was intentionally extracted from fixed Pauli measurements.
Nevertheless, the Pauli-axis diagnostics in Fig.~\ref{fig:deltaop_exp} establish $\delta_{\mathrm{ens}}>0$
operationally and provide an empirical constraint on how large preparation
variation could plausibly be on hardware. In particular, the relatively small
$\delta_{\mathrm{ens}}^{\min}$ required to account for the mitigated \texttt{ibm\_torino}
violations is not excluded by the observed nonstationarity, whereas for
\texttt{ibm\_fez} a local explanation would require substantially larger
$\delta_{\mathrm{ens}}$ that is not directly witnessed by the Pauli-axis
drift measurements. Hence, the CHSH data and the independent drift diagnostic
together delineate a preparation-dependent loophole: $|S|>2$ can coexist with
a breakdown of the stationarity premise required for the ideal CHSH bound.

It is instructive to compare the scale of preparation nonstationarity required
here with the degree of measurement dependence invoked in Hall-type analyses.
Hall showed that reproducing the quantum value $|S|=2\sqrt{2}$ within a local
model requires measurement dependence at the level
$M \gtrsim 2(\sqrt{2}-1)/3 \approx 0.14$, corresponding to a $\sim14\%$
violation of measurement independence.
By contrast, the preparation-dependent relaxation considered here requires
$\delta_{\mathrm{ens}}^{\min}=(|S|-2)/6$, which corresponds to ensemble
divergences of order $\sim3$--$4\%$ for the mitigated \texttt{ibm\_torino}
violations and $\sim11$--$12\%$ for \texttt{ibm\_fez}.
Although $M$ and $\delta_{\mathrm{ens}}$ quantify logically distinct
assumption relaxations and are not directly comparable, this contrast is
informative: modest levels of preparation nonstationarity, well below the
measurement-dependence scale required in Hall’s model, are sufficient to relax
the Bell bound to the experimentally observed values.
Importantly, unlike $M$, preparation nonstationarity is directly witnessed
operationally on hardware through bin-resolved drift diagnostics, without
invoking correlations between hidden variables and measurement choices.

To relate $\delta_{\mathrm{op}}$ to the CHSH aggregation actually performed, one must
account for how the execution schedule samples time across different
contexts. We capture this effect through a schedule-exposure factor
$\delta_{\mathrm{sched}}$, which quantifies the degree of temporal separation between contexts induced by the schedule. Because $\delta_{\mathrm{sched}}$ controls how temporal drift is mapped into
context dependence, a conservative schedule-aware lower bound on the
ensemble nonstationarity entering the CHSH experiment is
\begin{equation}
\delta_{\mathrm{ens}} \;\gtrsim\; \delta_{\mathrm{sched}}\,\delta_{\mathrm{op}}.
\end{equation}
This bound is not an equality and does not purport to measure the total
ensemble variation; rather, it quantifies the minimum nonstationarity that
must be present in the aggregated Bell data given the observed drift and the
actual execution schedule.

For round-robin execution, $\delta_{\mathrm{sched}}=0$ by construction, so
the schedule-induced lower bound is zero even when
$\delta_{\mathrm{op}}>0$. This reflects the fact that round-robin sampling
suppresses the exposure of drift in Bell aggregation without eliminating the
drift itself. In contrast, for unbalanced schedules,
$\delta_{\mathrm{sched}}\approx 1$, yielding
$\delta_{\mathrm{ens}}\gtrsim\delta_{\mathrm{op}}$.
Using the experimentally observed values
$\delta_{\mathrm{op}}\approx 0.089$ for \texttt{ibm\_torino} and
$\delta_{\mathrm{op}}\approx 0.062$ for \texttt{ibm\_fez} under unbalanced
execution, this establishes nonzero ensemble nonstationarity entering the
corresponding CHSH tests. Using $\underline{\delta}_{\mathrm{ens}}\equiv \delta_{\mathrm{sched}}\delta_{\mathrm{op}}$
as a conservative schedule-aware lower bound, we may define a minimal
relaxed bound under a LHV model. 
\begin{equation}
S_{\mathrm{LHV}}^{\mathrm{min}} \;=\; 2 + 6\,\underline{\delta}_{\mathrm{ens}}
\;\le\; 2 + 6\,\delta_{\mathrm{ens}}.
\end{equation}
For unbalanced execution, this gives
$S_{\mathrm{LHV}}^{\mathrm{min}}\approx 2.53$ on \texttt{ibm\_torino}
and $2.37$ on \texttt{ibm\_fez}. On \texttt{ibm\_torino}, the mitigated CHSH value satisfies
$|S|<S_{\mathrm{LHV}}^{\min}$, indicating that even this conservative bound is
sufficient to accommodate the observed correlations within the preparation
nonstationarity loophole. In contrast, on \texttt{ibm\_fez} we observe
$|S|>S_{\mathrm{LHV}}^{\min}$ after mitigation, showing that the \emph{minimum}
relaxation implied by our bound is not, by itself, enough to account for the
measured violation. This does not rule out a larger $\delta_{\mathrm{ens}}$
during the CHSH acquisition (e.g., axis-dependent drift or additional
schedule-induced context separation), but it demonstrates that the loophole
strength required to explain the \texttt{ibm\_fez} data would have to exceed
our conservative lower bound. 

\subsection{Qiskit-Aer simulations using a deterministic LHV model}

To connect the experimentally observed single-channel nonstationarity
(Fig. \ref{fig:deltaop_exp}) to relaxed Bell bounds, we adapt a simple
finite hidden-variable model from the class of deterministic, no-signalling
models introduced by Hall~\cite{hall2010local}.
The modification is minimal: we preserve deterministic local response
functions and operational measurement independence, but allow the preparation
ensemble to vary slowly in time.

We take a discrete hidden-variable space
$\Lambda=\{\lambda_1,\ldots,\lambda_5\}$, where each $\lambda_j$ may be
interpreted as an effective hardware ``mode'' of a given qubit pair.
Local outcomes are deterministic,
$A(a,\lambda_j),B(b,\lambda_j)\in\{\pm1\}$, with response functions chosen as in
Tab.~\ref{tab:lambda_table}.
For a fixed preparation ensemble, this reproduces Hall’s finite model and
yields $S=2+6p$ for a suitable mixture parameter $p$.

\begin{table}[t]
\centering
\caption{Deterministic local response functions used in the finite hidden-variable model.
Overall sign flips of $A$ or $B$ leave $|S|$ invariant and are omitted.}
\label{tab:lambda_table}
\begin{tabular}{c|cccc}
\hline
$\lambda_j$ & $A(x,\lambda_j)$ & $A(x',\lambda_j)$ & $B(y,\lambda_j)$ & $B(y',\lambda_j)$ \\
\hline
$\lambda_1$ & $+1$ & $+1$ & $+1$ & $+1$ \\
$\lambda_2$ & $+1$ & $-1$ & $+1$ & $+1$ \\
$\lambda_3$ & $+1$ & $+1$ & $+1$ & $-1$ \\
$\lambda_4$ & $+1$ & $-1$ & $-1$ & $+1$ \\
$\lambda_5$ & $+1$ & $+1$ & $+1$ & $+1$ \\
\hline
\end{tabular}
\end{table}

To model experimentally relevant drift, we introduce a discrete time (or bin)
index $k$ and allow the preparation ensemble to vary as
$\pi_{ab}^{(k)}(\lambda_j)$ according to Tab.~\ref{tab:pi_table_time}, with
$0\le p_k\le 1/3$.
For each fixed bin $k$, the preparation ensemble is identical across all CHSH
contexts, so that measurement independence is preserved within each bin.
The Bell parameter in bin $k$ is therefore
$S^{(k)}=2+6p_k$, and temporal variation of $p_k$ induces bin-resolved changes
in correlators without introducing context dependence at the preparation level. The operational single-channel drift is
\begin{equation}
\delta_{\mathrm{op}}^{(ab)}
=\max_{k,k'} d_{\mathrm{TV}}\!\left(\pi_{ab}^{(k)},\pi_{ab}^{(k')}\right)
=|p_k-p_{k'}|,
\end{equation}
directly mirroring the experimentally observed bin-to-bin variation in outcome
statistics for a fixed measurement setting.
Thus, nonzero $\delta_{\mathrm{op}}$ reflects temporal preparation
nonstationarity rather than correlations between $\lambda$ and the measurement
choices.

\begin{table}[]
\centering
\caption{Time-resolved preparation ensembles $\pi^{(k)}_{ab}(\lambda_j)$ for each CHSH context $(a,b)$.
Here $0\le p_k\le 1/3$ for all bins $k$. For fixed $k$ this reduces to Hall’s finite model.}
\label{tab:pi_table_time}
\begin{tabular}{c|cccc}
\hline
 & $\pi_{xy}^{(k)}$ & $\pi_{xy'}^{(k)}$ & $\pi_{x'y}^{(k)}$ & $\pi_{x'y'}^{(k)}$ \\
\hline
$\lambda_1$ & $p_k$    & $p_k$    & $p_k$    & $0$      \\
$\lambda_2$ & $p_k$    & $p_k$    & $0$      & $p_k$    \\
$\lambda_3$ & $p_k$    & $0$      & $p_k$    & $p_k$    \\
$\lambda_4$ & $0$      & $p_k$    & $p_k$    & $p_k$    \\
$\lambda_5$ & $1-3p_k$ & $1-3p_k$ & $1-3p_k$ & $1-3p_k$ \\
\hline
\end{tabular}
\end{table}

In an actual CHSH experiment, different measurement contexts $(a,b)$ are
executed over (possibly distinct) subsets of time bins $\mathcal{K}_{ab}$.
The effective preparation ensemble entering the Bell inequality is therefore
the time-averaged mixture
\begin{equation}
\bar{\pi}_{ab}(\lambda_j)
=\frac{1}{|\mathcal{K}_{ab}|}\sum_{k\in\mathcal{K}_{ab}}
\pi_{ab}^{(k)}(\lambda_j),
\end{equation}
with effective parameters
$\bar{p}_{ab}=|\mathcal{K}_{ab}|^{-1}\sum_{k\in\mathcal{K}_{ab}} p_k$.
The ensemble divergence relevant for the relaxed CHSH bound is
\begin{equation}
\delta_{\mathrm{ens}}
=\max_{(ab),(a'b')} d_{\mathrm{TV}}\!\left(\bar{\pi}_{ab},\bar{\pi}_{a'b'}\right)
=\max_{(ab),(a'b')}|\bar{p}_{ab}-\bar{p}_{a'b'}|.
\end{equation}

Crucially, $\delta_{\mathrm{ens}}$ can become nonzero even though the underlying
mechanism is purely temporal drift within a single preparation channel.
Nonuniform scheduling maps this drift onto different CHSH contexts, producing
effective context-conditioned ensembles.
In this sense, the experimentally observed single-channel nonstationarity
$\delta_{\mathrm{op}}>0$ implies that the hidden variable $\lambda$ is
effectively time dependent, and scheduling converts this time dependence into
context dependence at the level relevant for Bell inequalities.

To quantify this mechanism, we simulate the above model under conditions
matched to the experiments.
We initialize all four CHSH contexts with identical ensembles in each bin,
eliminating intrinsic measurement dependence.
Temporal drift is introduced via a linear ramp
$p_k\in[0,0.15]$ over $K=12$ bins, comparable to the variations inferred from
hardware.
Monte Carlo null tests with constant $p_k$ yield
$|S|\approx2.00$ with small $\delta_{\mathrm{op}}$ and $\delta_{\mathrm{ens}}$
consistent with finite-shot fluctuations.

\begin{table}[]
\centering
\caption{Simulation results for the drifting finite model with initially context-independent ensembles
($p_k\in[0,0.15]$, $K=12$ bins, 1024 shots/bin).
Balanced uses round-robin bin assignment; Unbalanced assigns early bins to $XY/XY'$ and late bins to $X'Y/X'Y'$.
The IID null uses constant $p_k=0.08$. Values are simulation averages.}
\label{tab:simulation_results}
\begin{tabular}{lccc}
\hline
Schedule & $S_r$ & $\delta_{\mathrm{op}}$ (global) & $\delta_{\mathrm{ens}}$ \\
\hline
Balanced   & 2.090 & 0.334 & 0.124 \\
Unbalanced & 2.191 & 0.169 & 0.288 \\
IID null   & 2.000 & 0.013 & 0.007 \\
\hline
\end{tabular}
\end{table}

Under balanced (round-robin) scheduling, all contexts sample identical
time-averaged ensembles, so $\delta_{\mathrm{ens}}$ remains small despite a
large global $\delta_{\mathrm{op}}$.
By contrast, under unbalanced scheduling—where early bins preferentially
contribute to $(x,y)$ and $(x,y')$ and late bins to $(x',y)$ and $(x',y')$—
the same temporal drift produces a substantial ensemble mismatch
$\delta_{\mathrm{ens}}\sim0.3$, accompanied by a relaxed Bell parameter
$|S|\gtrsim2.2$. These values are quantitatively consistent with the mitigated experimental results for \texttt{ibm\_torino} and \texttt{ibm\_fez}. These simulations validate a preparation-dependent loophole on NISQ hardware. Nonzero $\delta_{\mathrm{op}}$ provides an operational witness of temporal nonstationarity, while scheduling amplifies this drift into effective ensemble divergences $\delta_{\mathrm{ens}}$ sufficient to relax the CHSH bound. Unlike Hall’s original construction, no explicit correlation between hidden variables and measurement settings is introduced; nevertheless, the resulting context-conditioned ensembles occupy the same regime of relaxed Bell inequalities.

\section{Conclusion}

We have investigated the role of preparation nonstationarity in Bell--CHSH
experiments on superconducting quantum processors.
By relaxing the assumption of a single stationary preparation ensemble—while
preserving locality and measurement independence—we derived a relaxed Bell
inequality, $|S|\le 2+6\delta_{\mathrm{ens}}$, that quantifies how ensemble
divergence alone can inflate the Bell parameter.
This framework isolates a preparation-dependent loophole that is logically
distinct from Hall-type measurement dependence and is directly motivated by
known non-i.i.d.\ effects in superconducting hardware.

Operationally, we introduced $\delta_{\mathrm{op}}$ as a witness of preparation
nonstationarity, based on bin-resolved outcome statistics for fixed measurement
channels.
Pauli-axis measurements on \texttt{ibm\_fez} and \texttt{ibm\_torino} reveal
statistically significant drift exceeding matched Monte Carlo nulls, persisting
after full two-qubit readout mitigation.
These observations certify $\delta_{\mathrm{ens}}>0$ without relying on
assumptions about hidden-variable dynamics or measurement dependence.
In contrast, apparent drift extracted from CHSH-optimal settings is removed by
mitigation, confirming that such measurements conflate preparation drift with
basis-dependent measurement instability and are unsuitable for diagnosing
nonstationarity.

Although $\delta_{\mathrm{ens}}$ cannot be inferred directly from operational
data, the observed Bell violations imply a minimal ensemble divergence
$\delta_{\mathrm{ens}}^{\mathrm{req}}=(|S|-2)/6$ for any local explanation.
For the qubit pairs studied here, the required divergence ranges from a few to
$\sim10\%$, comparable to the scale of measurement dependence invoked in
Hall-type models.
Simulations of a drifting finite hidden-variable model demonstrate that purely
temporal preparation drift, when combined with nonuniform scheduling, can
amplify intra-context nonstationarity into effective context-dependent ensembles
sufficient to relax the CHSH bound without signaling.

Taken together, these results show that Bell violations on noisy
intermediate-scale quantum hardware can coexist with a breakdown of the
preparation-stationarity premise underlying the ideal CHSH bound.
This preparation-dependent loophole underscores the need for drift-aware
experimental protocols and operational diagnostics when using Bell tests for
quantum certification on real devices.

\textit{Acknowledgments} G.S. and R. P. thank Global Initiative of Academic Networks (GIAN) and IIT Delhi, which provided the environment in which this project was conceived. G.S. acknowledges ANRF-SERB Core Research Grant CRG/2023/005628.

\bibliographystyle{apsrev4-2}  
\bibliography{refs}            

@book{NielsenChuang,
  title={Quantum Computation and Quantum Information},
  author={Nielsen, Michael A. and Chuang, Isaac L.},
  publisher={Cambridge University Press},
  year={2010}
}

@article{Koch2007,
  title={Charge-insensitive qubit design derived from the Cooper pair box},
  author={Koch, Jens and Yu, Terri M. and Gambetta, Jay and Houck, Andrew A. and Schuster, David I. and Majer, Johannes and Blais, Alexandre and Devoret, Michel H. and Girvin, S. M. and Schoelkopf, R. J.},
  journal={Phys. Rev. A},
  volume={76},
  pages={042319},
  year={2007}
}

@article{javadi2024quantum,
  title={Quantum computing with Qiskit},
  author={Javadi-Abhari, Ali and Treinish, Matthew and Krsulich, Kevin and Wood, Christopher J and Lishman, Jake and Gacon, Julien and Martiel, Simon and Nation, Paul D and Bishop, Lev S and Cross, Andrew W and others},
  journal={arXiv preprint arXiv:2405.08810},
  year={2024}
}

@article{Chow2014,
  title={Implementing a Strand of a Scalable Fault-Tolerant Quantum Computing Fabric},
  author={Chow, J. M. and Gambetta, J. M. and C{\'o}rcoles, A. D. and Srinivasan, S. J. and Smolin, J. A. and Merkel, S. and Rozen, J. R. and Keefe, G. A. and Rothwell, M. B. and Ketchen, M. B. and Steffen, M.},
  journal={Nature},
  volume={508},
  pages={500--503},
  year={2014}
}

@article{ClauserHorne1974,
  title={Experimental consequences of objective local theories},
  author={Clauser, John F. and Horne, Michael A.},
  journal={Phys. Rev. D},
  volume={10},
  pages={526--535},
  year={1974}
}

@article{budroni2022kochen,
  title={Kochen-specker contextuality},
  author={Budroni, Costantino and Cabello, Ad{\'a}n and G{\"u}hne, Otfried and Kleinmann, Matthias and Larsson, Jan-{\AA}ke},
  journal={Reviews of Modern Physics},
  volume={94},
  number={4},
  pages={045007},
  year={2022},
  publisher={APS}
}

@article{howard2014contextuality,
  title={Contextuality supplies the ‘magic’for quantum computation},
  author={Howard, Mark and Wallman, Joel and Veitch, Victor and Emerson, Joseph},
  journal={Nature},
  volume={510},
  number={7505},
  pages={351--355},
  year={2014},
  publisher={Nature Publishing Group UK London}
}

@article{Arute2019,
  title={Quantum supremacy using a programmable superconducting processor},
  author={Arute, F. and Arya, K. and Babbush, R. and Bacon, D. and Bardin, J. C. and Barends, R. and Biswas, R. and Boixo, S. and Brandao, F. G. S. L. and Buell, D. A. and Burkett, B. and others},
  journal={Nature},
  volume={574},
  pages={505--510},
  year={2019}
}

@article{Neill2018,
  title={A Blueprint for Demonstrating Quantum Supremacy with Superconducting Qubits},
  author={Neill, Charles and Roushan, Pedram and Kechedzhi, Kevin and Boixo, Sergio and Mezzacapo, Antonio and Burkett, Brian and others},
  journal={Science},
  volume={360},
  pages={195--199},
  year={2018}
}

@article{Barends2014,
  title={Superconducting quantum circuits at the surface code threshold},
  author={Barends, R. and Kelly, J. and Megrant, A. and Veitia, A. and Sank, D. and others},
  journal={Nature},
  volume={508},
  pages={500--503},
  year={2014}
}

@article{Brunner2014,
author = {Brunner, Nicolas and Cavalcanti, Daniel and Pironio, Stefano and Scarani, Valerio and Wehner, Stephanie},
title = {Bell nonlocality},
journal = {Rev. Mod. Phys.},
volume = {86},
pages = {419--478},
year = {2014},
doi = {10.1103/RevModPhys.86.419}
}

@article{Bravyi2021,
author = {Bravyi, Sergei and Sheldon, Sarah and Kandala, Abhinav and McKay, David C. and Gambetta, Jay M.},
title = {Mitigating measurement errors in multiqubit experiments},
journal = {Phys. Rev. A},
volume = {103},
pages = {042605},
year = {2021},
doi = {10.1103/PhysRevA.103.042605}
}

@article{Proctor2020,
author = {Proctor, Timothy J. and Revelle, Melissa C. and Nielsen, Erik and Rudinger, Kenneth and Lobser, Daniel and Maunz, Peter and Blume-Kohout, Robin},
title = {Detecting and tracking drift in quantum information processors},
journal = {Nat. Commun.},
volume = {11},
pages = {5396},
year = {2020},
doi = {10.1038/s41467-020-19074-4}
}

@article{Setiawan2025,
author = {Setiawan, F. and Gramolin, Alexander V. and Matekole, Elisha S. and Krovi, Hari and Taylor, Jacob M.},
title = {Accurate and honest approximation of correlated qubit noise},
journal = {Quantum},
volume = {9},
pages = {1701},
year = {2025},
doi = {10.22331/q-2025-04-09-1701}
}

@article{Clauser1969,
author = {Clauser, John F. and Horne, Michael A. and Shimony, Abner and Holt, Richard A.},
title = {Proposed Experiment to Test Local Hidden-Variable Theories},
journal = {Phys. Rev. Lett.},
volume = {23},
pages = {880--884},
year = {1969},
doi = {10.1103/PhysRevLett.23.880}
}

@article{Bell1964,
author = {Bell, John S.},
title = {On the Einstein Podolsky Rosen Paradox},
journal = {Physics (Long Island City, N.Y.)},
volume = {1},
pages = {195--200},
year = {1964}
}

@article{Larsson2014,
author = {Larsson, Jan-{\AA}ke},
title = {Loopholes in Bell inequality tests of local realism},
journal = {J. Phys. A: Math. Theor.},
volume = {47},
pages = {424003},
year = {2014},
doi = {10.1088/1751-8113/47/42/424003}
}

@article{Ansmann2009,
author = {Ansmann, M. and Wang, H. and Bialczak, R. C. and Hofheinz, M. and Lucero, E. and Neeley, M. and O'Connell, A. D. and Sank, D. and Weides, M. and Wenner, J. and Cleland, A. N. and Martinis, John M.},
title = {Violation of {B}ell's inequality in Josephson phase qubits},
journal = {Nature},
volume = {461},
pages = {504--506},
year = {2009},
doi = {10.1038/nature08363}
}

@article{Storz2023,
author = {Storz, Simon and Sch{"a}r, Josua and Kulikov, Anatoly and Magnard, Paul and Kurpiers, Philipp et al.},
title = {Loophole-free {B}ell inequality violation with superconducting circuits},
journal = {Nature},
volume = {617},
pages = {265--270},
year = {2023},
doi = {10.1038/s41586-023-05885-0}
}

@article{Kjaergaard2020,
author = {Kjaergaard, Morten and Schwartz, Mollie E. and Braum{"u}ller, Jochen and Krantz, Philip and Wang, Joel~I.-J. and Gustavsson, Simon and Oliver, William D.},
title = {Superconducting qubits: current state of play},
journal = {Annu. Rev. Condens. Matter Phys.},
volume = {11},
pages = {369--395},
year = {2020},
doi = {10.1146/annurev-conmatphys-031119-050605}
}

@article{Kim2023,
author = {Kim, Youngseok and Nawrocki, Eric P. and Chen, Jeng-Bang and Choi, Joonhee et al.},
title = {Evidence for the utility of quantum computing before fault tolerance},
journal = {Nature},
volume = {618},
pages = {500--505},
year = {2023},
doi = {10.1038/s41586-023-06096-3}
}

@article{wenner2011surface,
  title={Surface loss simulations of superconducting coplanar waveguide resonators},
  author={Wenner, James and Barends, R and Bialczak, RC and Chen, Yu and Kelly, J and Lucero, Erik and Mariantoni, Matteo and Megrant, A and O’Malley, PJJ and Sank, D and others},
  journal={Applied Physics Letters},
  volume={99},
  number={11},
  year={2011},
  publisher={AIP Publishing}
}

@article{livingston2022experimental,
  title={Experimental demonstration of continuous quantum error correction},
  author={Livingston, William P and Blok, Machiel S and Flurin, Emmanuel and Dressel, Justin and Jordan, Andrew N and Siddiqi, Irfan},
  journal={Nature communications},
  volume={13},
  number={1},
  pages={2307},
  year={2022},
  publisher={Nature Publishing Group UK London}
}

@article{kelly2015state,
  title={State preservation by repetitive error detection in a superconducting quantum circuit},
  author={Kelly, Julian and Barends, Rami and Fowler, Austin G and Megrant, Anthony and Jeffrey, Evan and White, Theodore C and Sank, Daniel and Mutus, Josh Y and Campbell, Brooks and Chen, Yu and others},
  journal={Nature},
  volume={519},
  number={7541},
  pages={66--69},
  year={2015},
  publisher={Nature Publishing Group UK London}
}

@article{hutchings2017tunable,
  title={Tunable superconducting qubits with flux-independent coherence},
  author={Hutchings, MD and Hertzberg, Jared B and Liu, Yebin and Bronn, Nicholas T and Keefe, George A and Brink, Markus and Chow, Jerry M and Plourde, BLT},
  journal={Physical Review Applied},
  volume={8},
  number={4},
  pages={044003},
  year={2017},
  publisher={APS}
}

@article{harper2020efficient,
  title={Efficient learning of quantum noise},
  author={Harper, Robin and Flammia, Steven T and Wallman, Joel J},
  journal={Nature Physics},
  volume={16},
  number={12},
  pages={1184--1188},
  year={2020},
  publisher={Nature Publishing Group UK London}
}

@article{zhao2023mitigation,
  title={Mitigation of quantum crosstalk in cross-resonance-based qubit architectures},
  author={Zhao, Peng},
  journal={Physical Review Applied},
  volume={20},
  number={5},
  pages={054033},
  year={2023},
  publisher={APS}
}

@article{rudinger2019probing,
  title={Probing context-dependent errors in quantum processors},
  author={Rudinger, Kenneth and Proctor, Timothy and Langharst, Dylan and Sarovar, Mohan and Young, Kevin and Blume-Kohout, Robin},
  journal={Physical Review X},
  volume={9},
  number={2},
  pages={021045},
  year={2019},
  publisher={APS}
}

@article{goswami2021experimental,
  title={Experimental characterization of a non-Markovian quantum process},
  author={Goswami, K and Giarmatzi, C and Monterola, C and Shrapnel, S and Romero, J and Costa, F},
  journal={Physical Review A},
  volume={104},
  number={2},
  pages={022432},
  year={2021},
  publisher={APS}
}

@article{filenga2020non,
  title={Non-Markovian memory in a measurement-based quantum computer},
  author={Filenga, D and Mahlow, F and Fanchini, FF},
  journal={Physical Review A},
  volume={102},
  number={4},
  pages={042615},
  year={2020},
  publisher={APS}
}

@article{mcewen2021removing,
  title={Removing leakage-induced correlated errors in superconducting quantum error correction},
  author={McEwen, Matt and Kafri, Dvir and Chen, Z and Atalaya, Juan and Satzinger, KJ and Quintana, Chris and Klimov, Paul Victor and Sank, Daniel and Gidney, C and Fowler, AG and others},
  journal={Nature communications},
  volume={12},
  number={1},
  pages={1761},
  year={2021},
  publisher={Nature Publishing Group UK London}
}

@article{hall2010local,
  title={Local Deterministic Model of Singlet State Correlations Based on Relaxing<? format?> Measurement Independence},
  author={Hall, Michael JW},
  journal={Physical review letters},
  volume={105},
  number={25},
  pages={250404},
  year={2010},
  publisher={APS}
}

@article{barrett2011much,
  title={How much measurement independence is needed to demonstrate nonlocality?},
  author={Barrett, Jonathan and Gisin, Nicolas},
  journal={Physical review letters},
  volume={106},
  number={10},
  pages={100406},
  year={2011},
  publisher={APS}
}

@article{friedman2019relaxed,
  title={Relaxed Bell inequalities with arbitrary measurement dependence for each observer},
  author={Friedman, Andrew S and Guth, Alan H and Hall, Michael JW and Kaiser, David I and Gallicchio, Jason},
  journal={Physical Review A},
  volume={99},
  number={1},
  pages={012121},
  year={2019},
  publisher={APS}
}

@article{li2023tight,
  title={Tight bound on tilted CHSH inequality with measurement dependence},
  author={Li, Runze and Li, Dandan and Huang, Wei and Xu, Bingjie and Gao, Fei},
  journal={Physica A: Statistical Mechanics and its Applications},
  volume={626},
  pages={129037},
  year={2023},
  publisher={Elsevier}
}

@article{thinh2013bell,
  title={Bell tests with min-entropy sources},
  author={Thinh, Le Phuc and Sheridan, Lana and Scarani, Valerio},
  journal={arXiv preprint arXiv:1304.3598},
  year={2013}
}

@article{bharti2022noisy,
  title={Noisy intermediate-scale quantum algorithms},
  author={Bharti, Kishor and Cervera-Lierta, Alba and Kyaw, Thi Ha and Haug, Tobias and Alperin-Lea, Sumner and Anand, Abhinav and Degroote, Matthias and Heimonen, Hermanni and Kottmann, Jakob S and Menke, Tim and others},
  journal={Reviews of Modern Physics},
  volume={94},
  number={1},
  pages={015004},
  year={2022},
  publisher={APS}
}

@article{tilly2022variational,
  title={The variational quantum eigensolver: a review of methods and best practices},
  author={Tilly, Jules and Chen, Hongxiang and Cao, Shuxiang and Picozzi, Dario and Setia, Kanav and Li, Ying and Grant, Edward and Wossnig, Leonard and Rungger, Ivan and Booth, George H and others},
  journal={Physics Reports},
  volume={986},
  pages={1--128},
  year={2022},
  publisher={Elsevier}
}

@article{fauseweh2024quantum,
  title={Quantum many-body simulations on digital quantum computers: State-of-the-art and future challenges},
  author={Fauseweh, Benedikt},
  journal={Nature Communications},
  volume={15},
  number={1},
  pages={2123},
  year={2024},
  publisher={Nature Publishing Group UK London}
}

@article{bauer2023quantum,
  title={Quantum simulation for high-energy physics},
  author={Bauer, Christian W and Davoudi, Zohreh and Balantekin, A Baha and Bhattacharya, Tanmoy and Carena, Marcela and De Jong, Wibe A and Draper, Patrick and El-Khadra, Aida and Gemelke, Nate and Hanada, Masanori and others},
  journal={PRX quantum},
  volume={4},
  number={2},
  pages={027001},
  year={2023},
  publisher={APS}
}

@article{scholl2021quantum,
  title={Quantum simulation of 2D antiferromagnets with hundreds of Rydberg atoms},
  author={Scholl, Pascal and Schuler, Michael and Williams, Hannah J and Eberharter, Alexander A and Barredo, Daniel and Schymik, Kai-Niklas and Lienhard, Vincent and Henry, Louis-Paul and Lang, Thomas C and Lahaye, Thierry and others},
  journal={Nature},
  volume={595},
  number={7866},
  pages={233--238},
  year={2021},
  publisher={Nature Publishing Group UK London}
}

@article{di2025benchmarking,
  title={Benchmarking the quality of multiplexed qubit readout beyond assignment fidelity},
  author={Di Giovanni, Andras and Aasen, Adrian Skasberg and Lisenfeld, J{\"u}rgen and G{\"a}rttner, Martin and Rotzinger, Hannes and Ustinov, Alexey V},
  journal={Physical Review Applied},
  volume={24},
  number={4},
  pages={044043},
  year={2025},
  publisher={APS}
}

@article{maciejewski2020mitigation,
  title={Mitigation of readout noise in near-term quantum devices by classical post-processing based on detector tomography},
  author={Maciejewski, Filip B and Zimbor{\'a}s, Zolt{\'a}n and Oszmaniec, Micha{\l}},
  journal={Quantum},
  volume={4},
  pages={257},
  year={2020},
  publisher={Verein zur F{\"o}rderung des Open Access Publizierens in den Quantenwissenschaften}
}

@article{tabia2025almost,
  title={Almost device-independent calibration beyond Born’s rule: Bell tests for cross-talk detection},
  author={Tabia, Gelo Noel and Shih, Alex Yueh-Ting and Zheng, Jin-Yuan and Liang, Yeong-Cherng},
  journal={Quantum Science and Technology},
  volume={10},
  pages={035056},
  year={2025}
}

@article{lima2011optimal,
  title={Optimal measurement bases for Bell-tests based on the CH-inequality},
  author={Lima, G and Inostroza, EB and Vianna, RO and Larsson, J-{\AA} and Saavedra, C},
  journal={arXiv preprint arXiv:1111.0822},
  year={2011}
}

@article{rosset2012imperfect,
  title={Imperfect measurement settings: Implications for quantum state tomography and entanglement witnesses},
  author={Rosset, Denis and Ferretti-Sch{\"o}bitz, Raphael and Bancal, Jean-Daniel and Gisin, Nicolas and Liang, Yeong-Cherng},
  journal={Physical Review A—Atomic, Molecular, and Optical Physics},
  volume={86},
  number={6},
  pages={062325},
  year={2012},
  publisher={APS}
}

@article{gill2023optimal,
  title={Optimal statistical analyses of Bell experiments},
  author={Gill, Richard D},
  journal={AppliedMath},
  volume={3},
  number={2},
  pages={446--460},
  year={2023},
  publisher={MDPI}
}

@article{Endo2018,
  title={Practical quantum error mitigation for near-future applications},
  author={Endo, Suguru and Benjamin, Simon C. and Li, Ying},
  journal={Phys. Rev. X},
  volume={8},
  pages={031027},
  year={2018}
}

@article{cosco2024bayesian,
  title={Bayesian mitigation of measurement errors in multi-qubit experiments},
  author={Cosco, F and Plastina, F and Gullo, N Lo},
  journal={arXiv preprint arXiv:2408.00869},
  year={2024}
}

@article{gao2021practical,
  title={Practical guide for building superconducting quantum devices},
  author={Gao, Yvonne Y and Rol, M Adriaan and Touzard, Steven and Wang, Chen},
  journal={PRX quantum},
  volume={2},
  number={4},
  pages={040202},
  year={2021},
  publisher={APS}
}

@article{ware2021experimental,
  title={Experimental Pauli-frame randomization on a superconducting qubit},
  author={Ware, Matthew and Ribeill, Guilhem and Riste, Diego and Ryan, Colm A and Johnson, Blake and Da Silva, Marcus P},
  journal={Physical Review A},
  volume={103},
  number={4},
  pages={042604},
  year={2021},
  publisher={APS}
}

\end{document}